\documentclass[review]{elsarticle}

\usepackage{hyperref}
\usepackage{amsmath,amssymb,amsfonts}
\usepackage{booktabs}
\usepackage{caption}
\usepackage{siunitx}
\usepackage{bm}
\usepackage{scalerel}
\usepackage{multirow}
\usepackage{adjustbox}
\usepackage{subfigure}
\usepackage[para,online,flushleft]{threeparttable}
\usepackage{array}
\usepackage{makecell}
\usepackage{graphicx}

\journal{Journal of \LaTeX\ Templates}

\begin{document}

\begin{frontmatter}

\title{Robust Segmentation of CPR-Induced Capnogram Using U-net: Overcoming Challenges with Deep Learning}



\author {Andoni Elola$^1$, Imanol Ania$^2$, Xabier Jaureguibeitia$^2$, Henry Wang$^3$, Michelle Nassal$^3$, Ahamed Idris$^4$, Elisabete Aramendi$^2$\\
\ \\
$^1$ Department of Electronic Technology, University of the Basque Country, Bilbao, Spain \\
$^2$ Department of Communications Engineering, University of the Basque Country, Bilbao, Spain.\\
$^3$ Department of Emergency Medicine, The Ohio State University, Columbus.\\
$^4$  Department of Emergency
Medicine, University of Texas Southwestern Medical Center, Dallas.\\
}
\begin{abstract}

\textit{Objective:} The accurate segmentation of capnograms during cardiopulmonary resuscitation (CPR) is essential for effective patient monitoring and advanced airway management. This study aims to develop a robust algorithm using a U-net architecture to segment capnograms into inhalation and non-inhalation phases, and to demonstrate its superiority over state-of-the-art (SoA) methods in the presence of CPR-induced artifacts.

\textit{Materials and methods:} A total of 24354 segments of one minute extracted from 1587 patients were used to train and evaluate the model. The proposed U-net architecture was tested using patient-wise 10-fold cross-validation. A set of five features was extracted for clustering analysis to evaluate the algorithm performance across different signal characteristics and contexts. The evaluation metrics included segmentation-level and ventilation-level metrics, including ventilation rate and end-tidal-CO$_2$ values.

\textit{Results:} The proposed U-net based algorithm achieved an F1-score of 98\% for segmentation and 96\% for ventilation detection, outperforming existing SoA methods by 4 points. The root mean square error for end-tidal-CO$_2$ and ventilation rate were 1.9 mmHg and 1.1 breaths per minute, respectively.

Detailed performance metrics highlighted the algorithm's robustness against CPR-induced interferences and low amplitude signals. Clustering analysis further demonstrated consistent performance across various signal characteristics.

\textit{Conclusion:} The proposed U-net based segmentation algorithm improves the accuracy of capnogram analysis during CPR. Its enhanced performance in detecting inhalation phases and ventilation events offers a reliable tool for clinical applications, potentially improving patient outcomes during cardiac arrest.

\end{abstract}


\end{frontmatter}


\section{Introduction}

Cardiac arrest remains one of the leading causes of death globally, affecting millions of individuals each year \cite{wong2019epidemiology,tsao2023heart}. Despite advances in medical technology and emergency response protocols, the survival rates remain alarmingly low, underscoring the need for improved diagnostic and monitoring techniques during cardiopulmonary resuscitation (CPR).  

Capnography is a critical tool in monitoring patients during CPR \cite{soar2020adult}. It is a non-invasive monitoring technique that continuously measures and displays the concentration of carbon dioxide (CO$_2$) throughout the respiratory cycle, providing real-time feedback on ventilation and perfusion. In resuscitation science, capnography is essential for assessing the effectiveness of chest compressions \cite{sheak2015quantitative}, detecting return of spontaneous circulation (ROSC) \cite{elola2019capnography,elola2020multimodal,nassal2024temporal}, and assisting on making informed clinical decisions \cite{paiva2018use}. 

The accurate segmentation of capnogram is crucial for extracting meaningful information and ensuring effective patient management. In cardiac arrest scenarios, the most common applications of precise segmentation include the accurate measurement of end-tidal carbon dioxide (EtCO$_2$), which is the concentration of CO$_2$ at the end of an exhaled breath and serves as a critical indicator of ventilation efficacy and perfusion status \cite{sandroni2018capnography}. Additionally, accurate segmentation facilitates the determination of inter-breath interval (IBI) or ventilation frequency, which is essential not only for providing critical data on the rate of respiratory cycles but also for ensuring adherence to resuscitation guidelines and avoiding hyperventilation. These parameters are particularly important during CPR, as they aid in the guidance of resuscitation efforts \cite{soar2020adult}.

Several algorithms have been proposed to delineate capnograms in hemodynamically stable patients \cite{herry2014segmentation, singh2018automatic}. However, segmenting capnograms during CPR presents unique challenges due to the presence of interferences from chest compressions and artifacts caused by irregular breathing patterns. These factors can lead to significant added noise and waveform variations in amplitude, complicating the accurate identification and analysis of the signal features essential for effective ventilation monitoring \cite{leturiondo2020chest}.

\begin{figure}[h]
    \centering
    \includegraphics[width=0.6\linewidth]{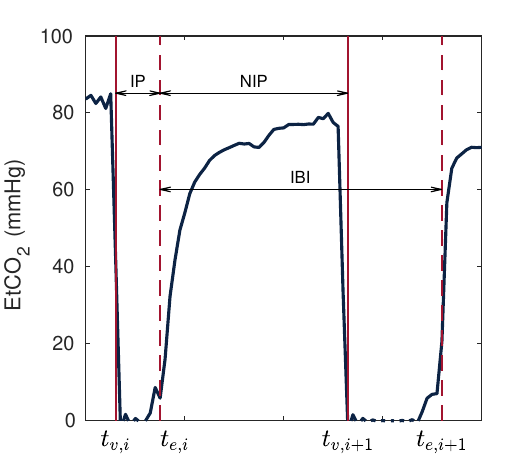}
    \caption{Capnography waveform during cardiac arrest, where the inhalation phase (IP), the non-inhalation phase (NIP) and the inter-breath interval (IBI) are shown for a ventilation event.}
    \label{fig:CO2_waveform}
\end{figure}

For ventilation rate computation Edelson et al. \cite{edelson2010capnography} proposed a novel algorithm to detect ventilation onsets, annotated as $t_v$ in Figure \ref{fig:CO2_waveform}. The algorithm relies on identifying abrupt decreases in the capnogram, which must then remain below a specific amplitude threshold for a certain period before increasing to higher plateau values. Subsequently, Aramendi et al. \cite{aramendi2017feasibility} developed an algorithm that detects expiratory onsets ($t_e$ in Figure \ref{fig:CO2_waveform}) by leveraging various capnogram amplitude and duration features and applying adaptive thresholds for detection of ventilatory events. Finally, the algorithm by Leturiondo et al. \cite{leturiondo2018influence} detects abrupt increases and decreases in the capnogram, utilizing thresholds based on amplitude and duration. While these state-of-the-art (SoA) algorithms can detect specific ventilatory events, such as ventilation and expiratory onsets, they do not explicitly address the segmentation of well known capnogram phases, which is highly required for capnography advanced monitoring during cardiac arrest. 

To address these challenges, we propose a novel approach utilizing a U-net architecture for the segmentation of capnograms during cardiac arrest. Our study aims to demonstrate the superiority of the U-net over classical signal processing techniques, specifically in the presence of artifacted and low amplitude capnograms. Additionally, we developed a clustering analysis to further evaluate the performance and robustness of our segmentation approach.

\section{Materials}

\begin{figure}[t!]
    \centering
    \subfigure[]{\includegraphics[width=0.6\linewidth]{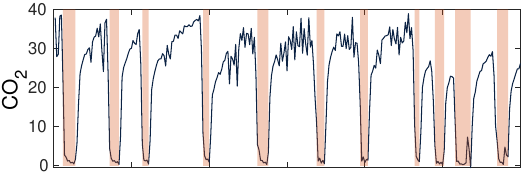}\label{fig:data_examples1}}
    \subfigure[]{\includegraphics[width=0.6\linewidth]{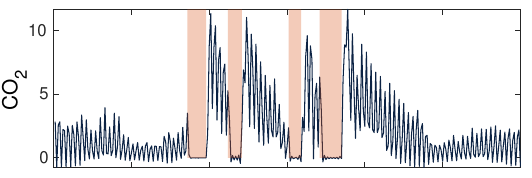}\label{fig:data_examples2}}
    \subfigure[]{\includegraphics[width=0.6\linewidth]{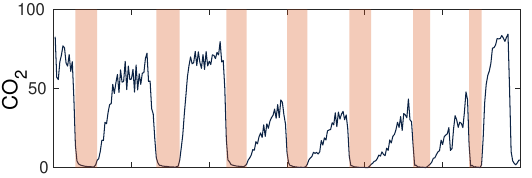}\label{fig:data_examples3}}
    \subfigure[]{\includegraphics[width=0.6\linewidth]{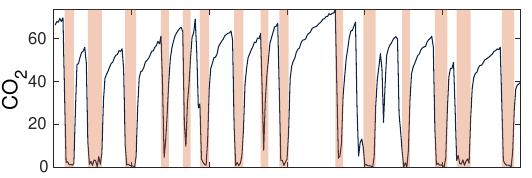}\label{fig:data_examples4}}
    \caption{Examples of 1-minute capnogram segments with the annotated inhalation phases shaded. High artifacts due to CPR compressions are observed as rapid fluctuations, especially in case b.}
    \label{fig:data_examples}
\end{figure}

The data used for this study were sourced from de-identified electronic defibrillator files of adult out-of-hospital cardiac arrest (OHCA) episodes enrolled in the Pragmatic Airway Resuscitation Trial (PART) \cite{wang2018effect}. Data were collected from three sites of the Resuscitation Outcome Consortium (ROC): Dallas-Fort Worth (TX), Milwaukee (WI), and Portland (OR). The episodes were recorded using HeartStart MRx (Philips Medical systems, Andover, MA) and ZOLL X-Series (ZOLL Medical, Chelmsford, MA) monitor-defibrillators. Capnography was acquired using sidestream configuration, with a sampling frequency of 125\,Hz and a resolution of 0.004\,mmHg in the MRx case, and with a sampling frequency of 20\,Hz and a resolution of 1\,mmHg in the X-Series case. Defibrillator files were converted to a common Matlab format (MathWorks Inc, Natick, MA) for further processing and analysis.

The inclusion criteria for this study required a minimum duration of 1 minute of capnogram for a patient to be considered. To be used as ground truth in the development of the algorithm, the inhalation phase (IP) of each ventilation was first automatically annotated in the capnogram using the algorithm proposed in \cite{aramendi2017feasibility} and then manually reviewed. The thoracic impedance (TI) was used to assist in this process when available in the defibrillator file. Capnogram intervals that were deemed overly uninterpretable were excluded from the study. Figure \ref{fig:data_examples} illustrates different examples of 1-minute capnogram segments, with the IP highlighted. They illustrate the wide variety of waveform patterns, some affected by CPR artifacts, which jeopardize the accuracy of automated methods for signal segmentation and analysis.

\section{Methods}

\begin{figure*}[t!]
    \centering
    \includegraphics[width=0.45\linewidth]{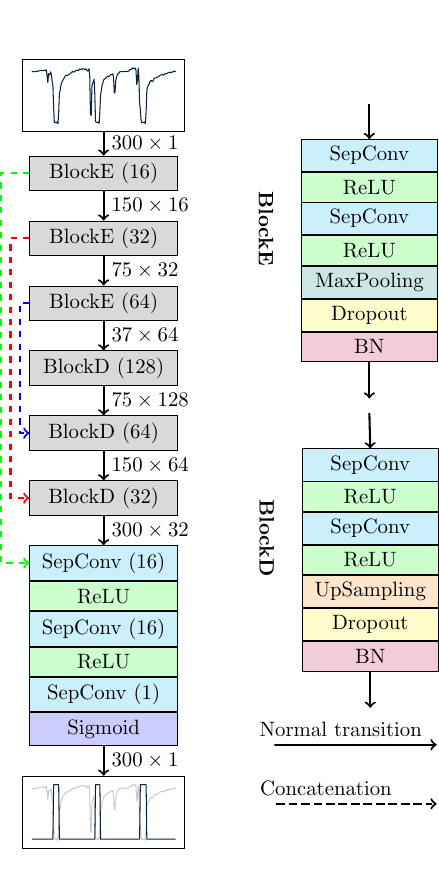}
    \caption{Structure of the U-Net architecture used for the segmentation of capnogram, where $s[n]$ is the input signal and $\hat{y}[n]$ the binary output signal obtained. For each block the number of filters used is indicated in parentheses.
    }
    \label{fig:cnn_scheme}
\end{figure*}

The proposed algorithm aimed to segment each ventilatory event of the capnogram into two different classes, thus discriminating the IP and the non-inhalation phase (NIP). Given a 1-minute capnogram \( s[n] \) of $N$ samples, the algorithm computes a binary signal \( \hat{y}[n] \), where \( \hat{y}[n] = 1 \) indicates the IP and \( \hat{y}[n] = 0 \) indicates the NIP, as shown in Figure \ref{fig:cnn_scheme}.

\subsection{Preprocessing}
The raw capnograms were first preprocessed to ensure uniformity and enhance signal quality. Initially, the signals were resampled to 5\,Hz to reduce the computational load on the U-Net architecture. Then, the signals were divided into 1-minute windows to reproduce the time frame used for ventilation feedback during CPR monitoring. Finally, the dataset was normalized to zero mean and unit variance, ensuring consistency and removing baseline variations. The mean and the variance were estimated using the whole training set.

The resulting dataset consisted in 24354 segments of 1 minute from 1587 patients. In terms of class balance, 23\% of signal samples corresponded to IPs ($\hat{y}[n]=1$) and the 77\% of the samples to NIPs ($\hat{y}[n]=0$).

\subsection{Capnography segmentation}

The core of our segmentation algorithm is a U-Net neural network \cite{ronneberger2015u}. The U-Net architecture, well known for its effectiveness in biomedical image segmentation, consists of an encoder-decoder structure with skip connections. 

The encoder compresses the input signal $s[n]$ into a lower-dimensional representation, while the decoder reconstructs the segmentation mask $\hat{y}[n]$. The full architecture of the network is shown in Figure \ref{fig:cnn_scheme}. Solid lines represent standard transitions between layers, with numbers alongside indicating the dimensions of the tensor. Dashed lines indicate concatenations between the output of one block and the input of another block. In this case, the encoder was made by 3 encoder blocks (BlockE) and 3 decoder blocks (blockD). The order of the filters was 6 for every separable convolutional layer; the number of filters is shown between parenthesis in Figure \ref{fig:cnn_scheme}. The max-pooling layers had a pooling size of 2 with a stride of 2. The network was trained on a dataset of annotated capnograms, using the 1-Dice coefficient as loss function; the optimization was carried out for 5 epochs, with a batch size of 32 (1-minute segments), and the employed optimizer was the standard Adam.

\subsection{Clustering} \label{sec:clustering}
Clustering was employed to identify distinct patterns within the segmented capnograms, which can be associated with different physiological conditions and types of interference. This unsupervised learning approach helps to categorize the signal segments into meaningful groups, thereby facilitating a more targeted analysis. By analyzing the performance of the algorithm across these different clusters, we can better understand its robustness and effectiveness in various scenarios, ensuring that the segmentation algorithm performs consistently well across a range of signal characteristics and conditions.

A set of five features, \(\mathbf{v}=\{v_1, v_2, \ldots, v_5\}\), was extracted from each segment to characterize its properties. These features were designed to capture the primary characteristics of the capnograms that challenge the accurate segmentation during cardiac arrest, specifically focusing on low EtCO\(_2\) values and interferences caused by CPR. 

The first feature, $v_1$, was the mean of the IBIs within the segment, computed as $t_{e,i+1}-t_{e,i}$ for each ventilation. The second feature, $v_2$, was the mean EtCO$_2$ value: for the \( i \)th ventilation, the maximum value of $s[n]$ was computed between $t_{e,i}$ and $t_{v,i+1}$, and the mean value among all ventilations was taken as $v_2$. The third feature, $v_3$, was the coefficient of variation, computed as the fraction between the standard deviation ($\sigma_s$) and the mean ($\mu_s$) of the $N$ samples of $s[n]$; this feature corresponds to higher values for high ventilation rates and/or highly distorted signals. The fourth feature, $v_4$, was the skewness of $s[n]$, computed as

\begin{equation}
    v_4 = \frac{1}{N} \sum_{i=1}^{N} \left(\frac{s[i] - \mu_s}{\sigma_s}\right)^3,
\end{equation}
and correlates with the relative duration of each phase. For those segments with IPs longer than NIPs (like the example shown in Figure \ref{fig:data_examples2} or \ref{fig:data_examples3}), $v_4$ is expected to be positive. For those cases with longer NIPs or plateaus as in the example shown in Figure \ref{fig:data_examples1} or\ref{fig:data_examples4}, $v_4$ should be negative. Finally, to compute the last feature, the power spectral density $|G(f)|$ of the segment $s[n]-\mu_s$ was first estimated using the periodogram, and $v_5$ was calculated as

\begin{equation}
    v_5=\frac{\int_{0}^{1} |G(f)| \,\mathrm{d}f}{\int_{0}^{2.5} |G(f)| \,\mathrm{d}f};
\end{equation}
this quantifies the proportion of spectral power concentrated in the 0-1\,Hz range. Current resuscitation guidelines recommend chest compression rates of 100-120 compressions per minute, meaning that the artifact associated with chest compressions typically appears above 1\,Hz. In contrast, the ventilation rate is generally much lower, with the primary frequency falling below 0.25\,Hz. Therefore, segments with less interference from chest compressions are expected to show higher values of $v_5$, with most power associated to the ventilation component of the capnogram.

The features were computed and standardized to zero mean and unit variance, and the K-means algorithm was selected for clustering due to its simplicity and effectiveness in partitioning data into distinct groups. We determined the optimal number of clusters $k=3$ based on the minimization of the Davies-Bouldin index, which measures the compactness and separation of the clusters \cite{davies1979cluster}.

\subsection{Evaluation}
For training and evaluating the model, a patient-wise 10-fold cross-validation approach was applied. To comprehensively assess the model's performance, both segmentation-level metrics and ventilation-level metrics were calculated.

To evaluate the segmentation power of the algorithm, the annotated mask and the predicted mask were compared. The following metrics were computed: sensitivity (Se), specificity (Sp), positive predictive value (PPV), negative predictive value (NPV), F1-score for the non-inspiratory (negative) class (\(\mathrm{F1^{(ni)}}\)), F1-score for the inspiratory (positive) class (\(\mathrm{F1^{(i)}}\)), and macro-averaged F1-score (\(\mathrm{F1^{(m)}}\)). 

To evaluate the power of the algorithm to detect ventilations, sensitivity (\(\mathrm{Se^{(v)}}\)), positive predictive value (\(\mathrm{PPV^{(v)}}\)), and F1-score (\(\mathrm{F1^{(v)}}\)) were computed. A ventilation was considered correctly detected (true positive) if the detected ventilation $t_{e,i}$ occurred within \(\pm0.5\) seconds of the manually annotated ground truth ventilation. For segments with \(\mathrm{Se^{(v)}}=0\) the F1$^\mathrm{(v)}$ was also set to zero.

Finally, errors in computing EtCO\(_2\) and ventilation rate were calculated. EtCO\(_2\) was determined as described in Section \ref{sec:clustering}, and ventilation rate was calculated as VR=$60/v_1$.

Each metric was computed for every segment, and the weighted mean and standard deviation across patients were reported.

\subsection{Comparison with SoA methods}

The algorithm was compared with three methods designed to detect ventilation or expiratory onsets \cite{edelson2010capnography,aramendi2017feasibility,leturiondo2018influence}. Although these methods were not specifically developed to segment the capnogram into IP and NIP phases, they all compute the instants $t_{v,i}$ and $t_{e,i}$. Using these instants, a mask was created as described earlier to facilitate comparison.

\section{Results}

\begin{table*}[t]
    \centering
    \begin{adjustbox}{max width=\linewidth}
    \begin{tabular}{ccccccccccccc} 

        \toprule
		&& \multicolumn{7}{c}{Segmentation Metrics} && \multicolumn{3}{c}{Ventilation Detection} \\
		\cmidrule(lr){2-9} \cmidrule(lr){10-13}
		&&  Sp & NPV  & F1$^{\mathrm{(ni)}}$ &  Se & PPV  & F1$^{\mathrm{(i)}}$ & F1$^{\mathrm{(m)}}$ && Se & PPV & F1$^{\mathrm{(v)}}$\\ 
		\cmidrule(lr){2-9} \cmidrule(lr){10-13}
		\vspace{-3mm}\\
		\multirow{3}{*} & Proposed\, & 0.98 (0.02) & 0.98 (0.02) & 0.98 (0.02) & 0.95 (0.05) & 0.94 (0.06) & 0.94 (0.05) & 0.96 (0.03)  && 0.97 (0.04) & 0.96 (0.06) & 0.96 (0.05)  \\
            & \cite{edelson2010capnography} & 0.98 (0.02) & 0.95 (0.04) & 0.96 (0.03) & 0.83 (0.11) & 0.94 (0.07) & 0.87 (0.09) & 0.92 (0.06)  && 0.90 (0.10) & 0.91 (0.10) & 0.89 (0.10)  \\
		& \cite{aramendi2017feasibility} & 0.95 (0.06) & 0.97 (0.03) & 0.95 (0.05) & 0.91 (0.10) & 0.89 (0.12) & 0.89 (0.11) & 0.92 (0.08)  && 0.90 (0.10) & 0.93 (0.09) & 0.91 (0.09)  \\
		& \cite{leturiondo2018influence} & 0.95 (0.06) & 0.96 (0.02) & 0.95 (0.04) & 0.87 (0.08) & 0.90 (0.12) & 0.87 (0.10) & 0.91 (0.07)  && 0.89 (0.11) & 0.94 (0.08) & 0.91 (0.09)  \\
		\vspace{-3mm}\\
		\bottomrule
    	\end{tabular}
    \end{adjustbox}
    \caption{Comparison of segmentation and ventilation detection metrics between the proposed U-net algorithm and state-of-the-art methods. All results are presented as patient-wise mean (standard deviation).}
    \label{tab:comp0}
\end{table*}


\subsection{Overall performance}
The proposed algorithm demonstrated outstanding performance, achieving an F1-score of 98\% for signal segmentation and 96\% for ventilation detection. These results highlight the algorithm's ability to accurately delineate and identify key features in capnograms during CPR. Table \ref{tab:comp0} provides a detailed comparison of our algorithm's performance against three SoA algorithms, showcasing significant improvements in both signal segmentation and ventilation detection metrics.

\subsection{Computing physiological variables}
\begin{figure}[t]
    \centering
    \subfigure[]{\includegraphics[width=0.47\linewidth]{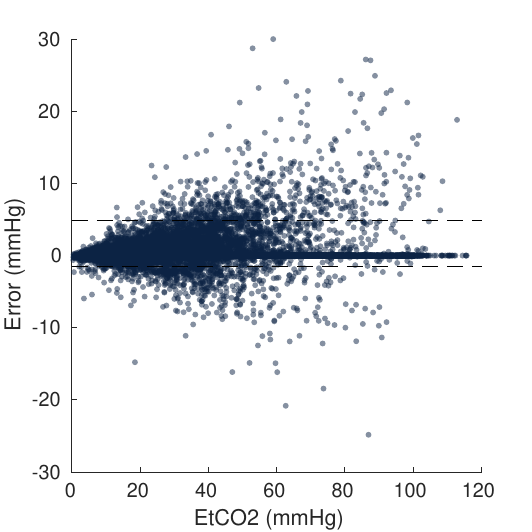}\label{fig:bland_altman_etco2}}
    \subfigure[]{\includegraphics[width=0.47\linewidth]{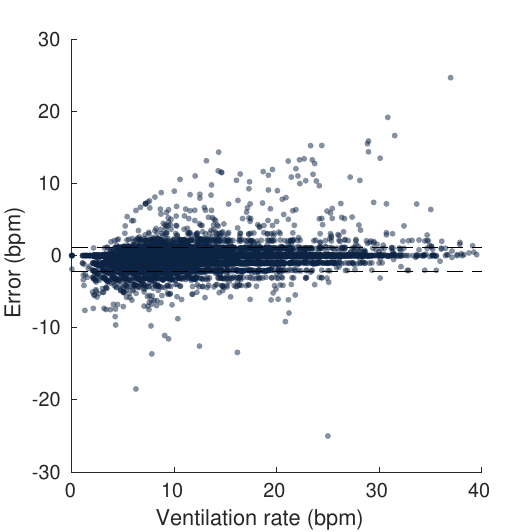}\label{fig:bland_altman_ventilation}}
    \caption{Bland-Altman plots for EtCO$_2$ in mmHG (panel a) and ventilation rate in breaths per minute (panel b). The mean difference (bias) and 95\% limits of agreement are shown}
    \label{fig:BA}
\end{figure}

To further validate the accuracy of proposed algorithm, we performed Bland-Altman analyses to assess the agreement between the U-net based measurements and the ground truth for both EtCO$_2$ and VR. Figure \ref{fig:bland_altman_etco2} shows the Bland-Altman plot for EtCO$_2$, and Figure \ref{fig:bland_altman_ventilation} shows the Bland-Altman plot for ventilation rate. The mean difference (bias) and the limits of agreement are indicated in both plots. Since the distribution of the error did not pass the Kolmogorov-Smirnov normality test, limits of agreement were estimated using percentiles of 97.5\% and 2.5\%.

 The mean difference for EtCO$_2$ was 0.3\,mmHg with 95\% limits of agreement ranging from -1.6 to 4.9\,mmHg, demonstrating a high level of accuracy and consistency with the reference values. The mean difference for VR was -0.14 breaths per minute (bpm) with 95\% limits of agreement from -2.2 to 1.1\,bpm, indicating reliable performance in detecting ventilation rates during CPR. These narrow intervals of confidence suggest that our algorithm provides precise estimations, crucial for effective patient monitoring during resuscitation.

When compared to SoA methods, our algorithm demonstrated the lowest error in computing both EtCO$_2$ and VR values The Root Mean Squared Error (RMSE) for EtCO$_2$ was 1.9\,mmHg and for VR it was 1.1\,bpm. The algorithm with the second-lowest error for EtCO$_2$ was Edelson et al. \cite{edelson2010capnography}, with an RMSE of 2.25\,mmHg. Computing the VR the algorithm proposed by Aramendi et al. \cite{aramendi2017feasibility} achieved the next best result with an RMSE of 2.3\,bpm.

\subsection{Clustering}

\begin{table}[h]
    \centering
    \begin{adjustbox}{max width=0.8\linewidth}
	\begin{tabular}{ccccccccccc} 
	    \toprule
		&& \multicolumn{2}{c}{Cluster1} && \multicolumn{2}{c}{Cluster2} && \multicolumn{2}{c}{Cluster3} \\
		\cmidrule(lr){3-4} \cmidrule(lr){6-7} \cmidrule(lr){9-10}
		&&  F1$^{\mathrm{(m)}}$ & F1$^{\mathrm{(v)}}$ &&  F1$^{\mathrm{(m)}}$ & F1$^{\mathrm{(v)}}$ &&  F1$^{\mathrm{(m)}}$ & F1$^{\mathrm{(v)}}$\\ 
        \midrule
		\vspace{-3mm}\\
		& Proposed\, & 0.97 (0.02)  & 0.97 (0.04) && 0.96 (0.02)  & 0.97 (0.02) && 0.93 (0.04)  & 0.93 (0.05)\\
            & \cite{edelson2010capnography} & 0.94 (0.04)  & 0.92 (0.07)  && 0.93 (0.03)  & 0.92 (0.04) && 0.88 (0.05)  & 0.84 (0.09) \\
		& \cite{aramendi2017feasibility} & 0.95 (0.05)  & 0.93 (0.07) && 0.94 (0.03)  & 0.93 (0.04) && 0.84 (0.08)  & 0.86 (0.07) \\
		& \cite{leturiondo2018influence} & 0.93 (0.04)  & 0.93 (0.07) && 0.92 (0.03)  & 0.91 (0.05) && 0.84 (0.07)  & 0.89 (0.06) \\	
		\vspace{-3mm}\\
		\bottomrule
	\end{tabular}
	\end{adjustbox}
    \caption{F1-scores for both segmentation and ventilation detection across the three clusters for the different algorithms}
    \label{tab:comp_clustering}
\end{table}

\begin{figure}[t]
    \centering
    \subfigure[$v_1$]{\includegraphics[width=0.3\linewidth]{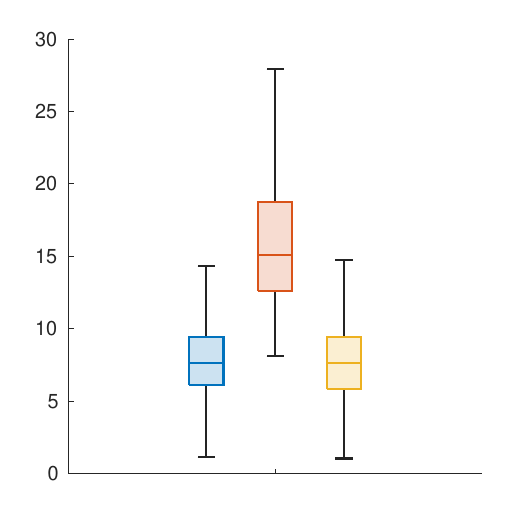}\label{fig:cluster_features1}}
    \subfigure[$v_2$]{\includegraphics[width=0.3\linewidth]{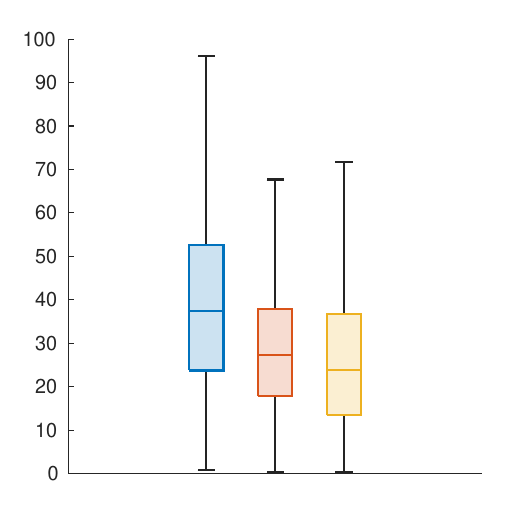}\label{fig:cluster_features2}}
    \subfigure[$v_3$]{\includegraphics[width=0.3\linewidth]{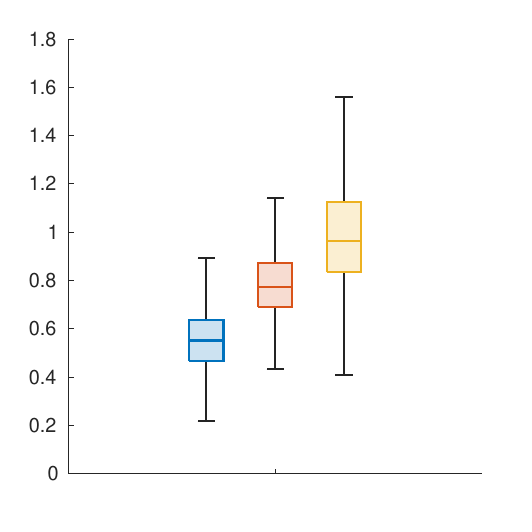}\label{fig:cluster_features3}}
    \subfigure[$v_4$]{\includegraphics[width=0.3\linewidth]{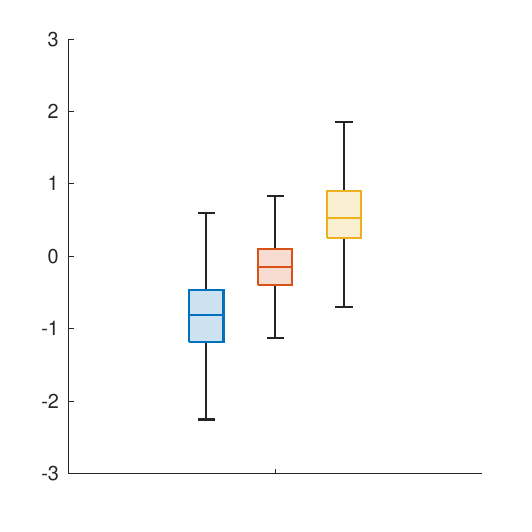}\label{fig:cluster_features4}}
    \subfigure[$v_5$]{\includegraphics[width=0.3\linewidth]{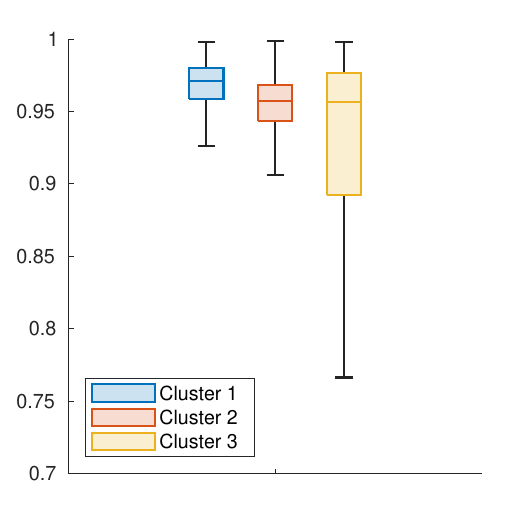}\label{fig:cluster_features5}}
    \caption{Boxplots illustrating the distribution of feature values across the three clusters. Each boxplot represents the statistical distribution of a specific feature for the three clusters identified in the study.}
    \label{fig:clustering_features}
\end{figure}

The unsupervised clustering based on the five features showed that from a total of 24354 segments, 13714 belong to Cluster 1, 5784 to Cluster 2, and 4856 to Cluster 3. Table \ref{tab:comp_clustering} presents the F1-scores for both segmentation and ventilation detection across these clusters and algorithms. Figure \ref{fig:clustering_features} shows the statistical distributions of each feature across the three clusters.

All algorithms consistently showed the highest performance for Cluster 1, indicating that it represents the most straightforward scenario for signal segmentation and analysis. This can be attributed to the moderate EtCO$_2$ values and relatively low interference levels, as shown by higher values of $v_2$, Figure \ref{fig:cluster_features2}, and $v_5$, Figure \ref{fig:cluster_features5}, compared to the values in other clusters. This condition facilitates the accurate segmentation and detection of ventilation events.

In Cluster 2, a slight decline in performance was observed for all algorithms. However, the most significant drop in performance was noted in Cluster 3 for all algorithms. This decline is more pronounced due to the combination of lower EtCO$_2$ amplitudes and increased interference levels, demonstrated by lower values of $v_2$ and $v_5$ in this cluster compared to the overall distributions. Some of the SoA methods are jeopardized by these conditions, leading to F1$^\mathrm{(m)}$/F1$^\mathrm{(v)}$ scores up to 9 points below our proposal.

Despite these challenges, the proposed algorithm consistently outperforms the SoA across all the three clusters. Its superior performance can be attributed to the deep learning-based approach, which effectively adapts to varying signal characteristics and maintains robustness even in clusters with higher levels of interference and lower amplitude signals.

The results of this clustering analysis highlight the importance of considering the variability of capnograms. The ability of the proposed algorithm to maintain high performance across diverse conditions underscores its potential applicability in real-world scenarios.

\subsection{Examples of ventilation detection}
\begin{figure*}[t]
    \centering
    \subfigure[]{\includegraphics[width=0.47\linewidth]{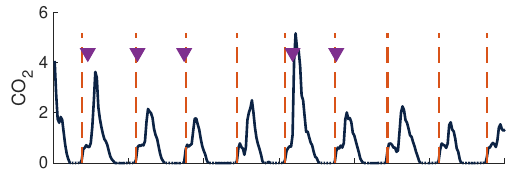}\label{fig:algorithm_examples1}}
    \subfigure[]{\includegraphics[width=0.47\linewidth]{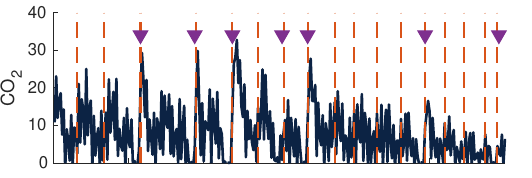}\label{fig:algorithm_examples2}}
    \subfigure[]{\includegraphics[width=0.47\linewidth]{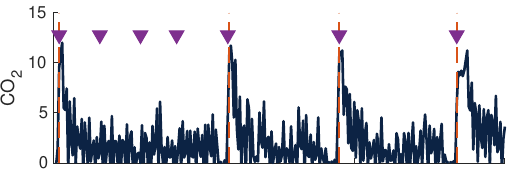}\label{fig:algorithm_examples3}}
    \subfigure[]{\includegraphics[width=0.47\linewidth]{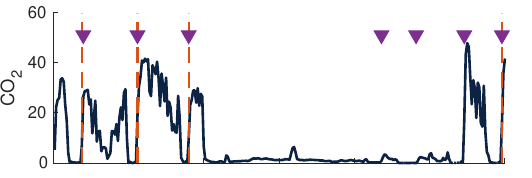}\label{fig:algorithm_examples4}}
    \subfigure[]{\includegraphics[width=0.47\linewidth]{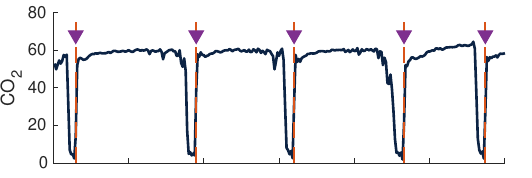}\label{fig:algorithm_examples5}}
    \subfigure[]{\includegraphics[width=0.47\linewidth]{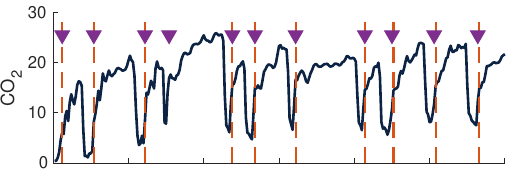}\label{fig:algorithm_examples6}}
    \subfigure[]{\includegraphics[width=0.47\linewidth]{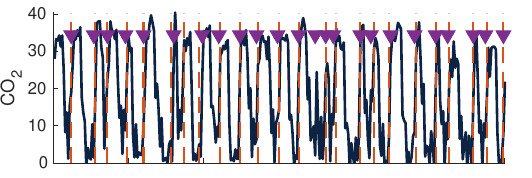}\label{fig:algorithm_examples7}}
    \subfigure[]{\includegraphics[width=0.47\linewidth]{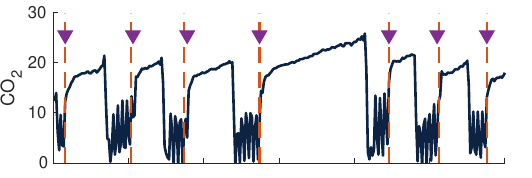}\label{fig:algorithm_examples8}}
    \caption{One-minute examples of capnograms, where dashed lines represent manually annotated ventilations (ground truth), and triangles indicate ventilations detected by the proposed algorithm.}
    \label{fig:algorithm_examples}
\end{figure*}

Figure \ref{fig:algorithm_examples} shows several examples of the performance of the algorithm in ventilation detection, where manual annotations are indicated by dashed lines, and detected ventilations by triangle markers. Some patterns remain challenging, as in Figure \ref{fig:algorithm_examples1}, which features a very low amplitude and a stepped NIP. Discriminating between real ventilations and artifacts can also be difficult. For instance, in Figure \ref{fig:algorithm_examples2}, the sensitivity (Se$^\mathrm{(v)}$) is low because the algorithm misinterpreted some waveform fluctuations as ventilations due to the presence of chest compressions. However, in Figures \ref{fig:algorithm_examples3} and \ref{fig:algorithm_examples4}, several false positives were detected, resulting in lower PPV$^\mathrm{(v)}$ scores. The last four panels of Figure \ref{fig:algorithm_examples} display successful examples where the F1$^\mathrm{(v)}$ of our algorithm exceeds 95\%. SoA methods \cite{edelson2010capnography,aramendi2017feasibility,leturiondo2018influence} performed worse with these examples (F1$^\mathrm{(v)} < 60\%$) 
due to factors such as short IPs (\ref{fig:algorithm_examples5}), baseline wandering (\ref{fig:algorithm_examples6}), high ventilation frequencies (\ref{fig:algorithm_examples7}), or artifacts in the inhalation phase (\ref{fig:algorithm_examples8}).

\section{Discussion}

In this manuscript, we propose an algorithm to segment capnograms into two distinct phases corresponding to inhalation and non-inhalation. This segmentation step is essential for accurately computing fundamental physiological variables such as VR and EtCO$_2$ during CPR. Additionally, it enables the analysis of more advanced features like the airway opening index (AOI), the relative duration of each phase, the slope of the plateau, or the number of notches within the plateau \cite{van1987computer,herry2014segmentation}. These measurements are key for assessing respiratory function and guiding clinical decisions during cardiac arrest.

When comparing the proposed algorithm with SoA methods, our deep learning based approach demonstrated significant superiority across all metrics and clusters. Furthermore, to the best of our knowledge, this study involved a cohort of patients (1587) larger than in any previous research on this topic, where the number of patients was limited to 37, 83, and 232 \cite{edelson2010capnography,aramendi2017feasibility,leturiondo2018influence}. This extensive dataset enhances the robustness and generalization of our findings, underscoring the potential of our algorithm to be accurately applied in diverse clinical scenarios.

The capnogram may contain respiratory-related information up to 10\,Hz \cite{yang2010new}, but in preliminary experiments we concluded that a sampling frequency of 5\,Hz was sufficient to delineate IPs and NIPs. Consequently, applying a low-pass filter with a cutoff frequency of 2.5\,Hz may be adequate to segment the capnogram during CPR. Lower sampling rates reduce the data volume, simplifying the network's structure by requiring fewer parameters, decreasing memory usage, and accelerating training time, all while preserving key information and improving segmentation performance by minimizing unnecessary high-frequency noise.

We selected the U-net architecture for its simplicity and effectiveness in segmenting medical images and signals \cite{aghalari2021brain,chen2023post,shu2024csca,pham2021ear,renna2019deep,krithika2022review}. As a fully convolutional network, the U-net excels at capturing spatial hierarchies within the data, making it well-suited for delineating the inhalation and non-inhalation phases of capnograms. Although we experimented with additional post-processing methods such as recurrent neural networks (RNNs), Hidden Markov Models (HMMs), and Conditional Random Fields (CRFs), these approaches did not yield significant improvement in segmentation accuracy. Thus, the U-net's ability to perform end-to-end learning and its proven success in similar applications made it the optimal choice for our segmentation task.

Despite the promising results obtained, it is important to acknowledge certain limitations of our study. Primarily, the algorithm was developed and evaluated using data from adult patients who experienced OHCA. This specific population may not accurately represent other patient groups, such as children or those in hospital settings with different medical conditions. Consequently, future studies should focus on evaluating the algorithm's effectiveness across a broader range of clinical scenarios and populations, including pediatric patients and situations where respiratory or cardiac conditions differ significantly from cardiac arrest. This approach will help determine the algorithm's applicability and effectiveness in more diverse contexts.

\section{Conclusion}

This study presents a novel approach for segmenting capnogram during CPR using a U-Net-based deep learning algorithm. Our results demonstrate that this method outperforms existing SoA techniques in segmenting the waveform and detecting ventilatory parameters. The algorithm not only achieved superior performance in terms of both EtCO$_2$ and VR measurement, but it also effectively managed the complexities and artifacts inherent to CPR scenarios.

The application of this algorithm for monitoring patients during CPR could introduce meaningful progress in the processing of capnogram and contribute to the advancement of emergency science.

\section{Acknowledgments}

This research has been partially supported by the MCIN/ AEI/10.13039/501100011033/, by FEDER Una manera de hacer Europa
through grant PID2021-122727OB-I00, and by the Basque Government through grant IT1717-22.

\bibliography{mybibfile}

\end{document}